# A mini review on the chemistry and catalysis of the water gas shift reaction

**Abstract:**

Bifunctional/bimetallic catalysts are a set of important catalytic materials that find their applications in many reaction systems.[1-11] Among them, water gas shift (WGS) reaction is a chemical reaction in which carbon monoxide reacts with water vapor to form carbon dioxide and hydrogen ($CO+H_2O \rightarrow CO_2+H_2$). It is an important reaction industrially used in conjunction with steam reforming of hydrocarbons for the production of high purity hydrogen. Grenoble *et al* examined the roles of both active metals and metal oxide support on the kinetics of the WGS reaction.[12] They found out that the turn over numbers of various $Al_2O_3$ supported transition metals decreased in the trend of Cu, Re, Co, Ru, Ni, Pt, Os, Au, Fe, Pd, Rh, and Ir, which corresponds nicely to the observed volcano shaped correlation between catalytic activities and respective CO adsorption heat. This is a strong indication that CO gets activated on the metal surface during the reaction and different metals have different activation energies. The authors also observed that the turn over number of $Pt/Al_2O_3$ was one order of magnitude higher than that of $Pt/SiO_2$, indicating a strong support effect, which the authors ascribed to the activation of $H_2O$. Based on the observed metal and support effects, the authors proposed a bifunctional reaction mechanism for $Al_2O_3$ supported metal catalysts by assuming CO chemisorbs onto metal sites non-dissociatively and $H_2O$ chemisorbs onto support metal oxide sites dissociatively (Scheme 1). Formic acid-surface complex was proposed to be an intermediate, and the exchange of formic acid from support sites onto metal sites was thought to be the rate limiting step. The Langmuir-Hinshelwood kinetics model from the proposed mechanism demonstrates that the partial pressure dependencies for CO and H₂O are close to $0^{th}$ and $0.5^{th}$ order for most metals, respectively, which are in consistent with the experimental parameters.

$$CO + M \underset{}{\overset{K_{CO}}{\longleftrightarrow}} CO-M$$

$$H_2O + 2S \underset{}{\overset{K_w}{\longleftrightarrow}} HO-S + H-S$$

$$CO-M + HO-S + H-S \overset{k_f}{\rightarrow} HCOOH-S + M$$

$$HCOOH-S + M \overset{k_{WGS}}{\longrightarrow} HCOOH-M + S \qquad \text{Rate limiting step}$$

$$HCOOH-M \overset{k_{fast}}{\longrightarrow} H_2 + CO_2 + M$$

**Scheme 1: Proposed reaction mechanism**

**An introduction to WGS reaction**

The water-gas shift (WGS) reaction (CO + H₂O = CO₂ + H₂) is used in industrial hydrogen production as well as an integral component of fuel processing for fuel cell applications. Understanding this reaction is thus critical for any practical device that generates $H_2$ from a hydrocarbon source. Its purpose is to produce hydrogen and to reduce the level of CO for final cleanup by preferential oxidation.

**A summary of the methodology**

All the catalysts were prepared by incipient wetness method. Aqueous solutions of metal salts were prepared and then mixed with solid support materials ($Al_2O_3$, $SiO_2$ and C) using different amounts depending on the loading of metals on support. The solids were then calcinated under $H_2$ flow at 500 °C to form nano-sized active metal domain on the support. All the as-synthesized catalysts were pretreated with $H_2$ flow at 500 °C for 1 hr. (for Au and Cu, 250 °C was used) prior to kinetic study to fully reduce oxidized metal surface.

Catalysts used in this study were characterized by $H_2$ and CO chemisorption to measure metal dispersion. X-ray line broadening technique was used for Au and Cu catalysts, as both of them have very weak $H_2$ and CO absorption.

The reaction was carried out in a differential reactor with constant feed pressures (CO 24.3 kPa, $H_2O$ 31.3 kPa), but with varied temperatures from 130 °C to 380 °C to maintain CO conversion of less than 5%, assuring negligible mass and heat transfer. Gas Chromatography (GC) was used to analyze the inlet and outlet stream composition. The catalytic activity (turnover number) reported in the study was corrected to 300 °C using Arrhenius Equation for all the catalysts.

**Effect of active metal**

Turn over numbers of various $Al_2O_3$ supported transition metals decreased in the trend of Cu, Re, Co, Ru, Ni, Pt, Os, Au, Fe, Pd, Rh, and Ir. The experimental orders of reaction with respect to $H_2O$ and CO were from near zero to 0.8 and from -0.4 to +0.6, respectively.

The authors plotted the logarithm of turnover rate against periodic group number for various metals, and found out that VIII metals have comparable activities, with the most active metal Ru being only 60 times more active than the least active metal Ir. However, for metals outside VIII group, such as Cu, is 50 times more active than Ru, and nearly 4000 more active than Ir.

Furthermore, for noble metals, there is a minimum in activity at the Group VIII$_2$, Rh and Ir. The periodic trends observed for the iron triad metals (Fe, Co and Ni) are opposite to noble metals, with Co the most active instead of least.

The authors ascribed this observed periodic trends to the ability of CO chemisorption on the metal surface reflected by the heat of CO adsorption (Figure 1).

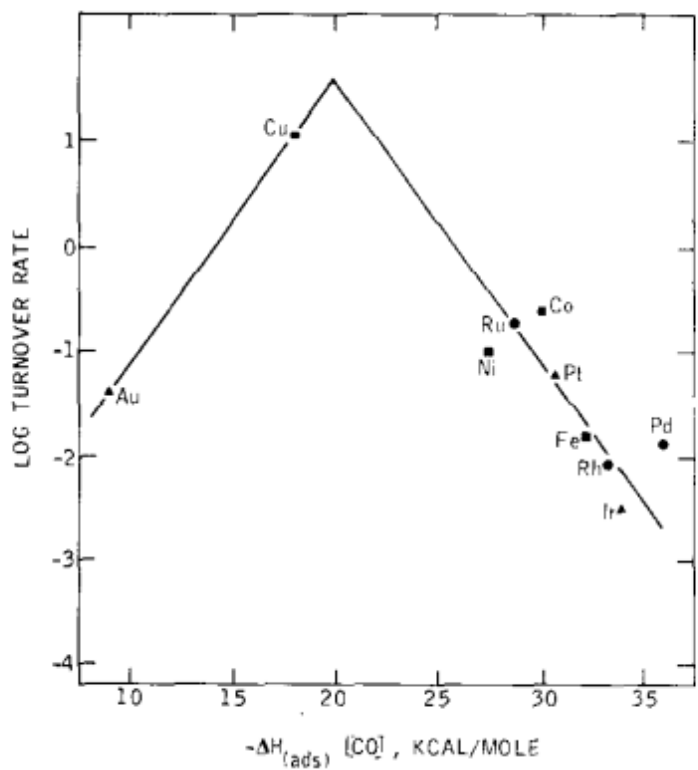

**Figure 1. Volcano-shaped relationship between metal turnover number and heat of adsorption of CO (Adapted from [12])**

They argued that the strength of metal-CO interaction should neither be very strong such that CO is hard to detach from the metal surface, nor be very weak such that the concentration of CO-M species is low for next reaction step. Cu has a moderate interaction as inflected by its position near the peak in Figure 1. That is why Cu is the most active metal investigated. In comparison, Au has a too weak interaction, while other metals such as Pd, Ru and Ir have too strong interactions, which make all of them less active than Cu.

**Effect of support**

A Comparison of the Kinetic Parameters of Alumina- and Nonalumina-Supported Metals

| Catalyst | $T^a$ | $b$ | $c$ | $E^d$ | $A^e$ | $N^f$ | Relative activity |
|---|---|---|---|---|---|---|---|
| Pt/Al$_2$O$_3$ | 270 | −0.21 ± 0.03 | 0.75 ± 0.04 | 19.6 ± 1.3 | 1.90 × 10$^6$ | 0.0635 | 90 |
| Pt/SiO$_2$ | 340 | −0.08 ± 0.05 | 0.69 ± 0.08 | 19.1 ± 0.8 | 1.18 × 10$^5$ | 0.0061 | 9 |
| Pt/C | 340 | 0.13 ± 0.05 | 0.35 ± 0.17 | 25.5 ± 1.4 | 3.84 × 10$^6$ | 0.0007 | 1 |
| Rh/Al$_2$O$_3$ | 330 | −0.10 ± 0.02 | 0.44 ± 0.02 | 23.0 ± 1.3 | 5.10 × 10$^6$ | 0.0086 | 13 |
| Rh/SiO$_2$ | 350 | −0.24 ± 0.03 | 0.53 ± 0.12 | 22.8 ± 2.5 | 3.23 × 10$^5$ | 0.0007 | 1 |

$^a$ Temperature at which reaction orders were determined, °C.
$^b$ Order with respect to CO.
$^c$ Order with respect to H$_2$O.
$^d$ Apparent activation energy, kcal/mole.
$^e$ Pre-exponential factor, molecules/sec/metal site in the equation $N = A\ exp(-E/RT)$.
$^f$ Turnover rate after 1 hr on stream and conditions of 300°C, $P_{CO}$ = 24.3 kPa, $P_{H_2O}$ = 31.4 kPa.

**Table 1. A comparison of the kinetic parameters of catalysts with various supports (Adapted from [12])**

The authors have also investigated the effect of support by depositing Pt and Rh on various support materials such as Al$_2$O$_3$, SiO$_2$, and C. They observed that the turn over numbers of Pt/Al$_2$O$_3$ and Rh/Al$_2$O$_3$ were both one order of magnitude higher than their SiO$_2$ supported counterparts (Table 1), indicating a strong support effect, which the authors ascribed to the activation of H$_2$O.

Figure 2 shows two possible ways of H$_2$O activation on Al$_2$O$_3$-dissociative adsorption and nondissociative adsorption, respectively. For the nondissociative way, water gets chemisorbed at the Lewis acid centers of Al$_2$O$_3$. Since high temperature was used in this reaction, the authors argued that both activation mechanism exist while nondissociative one is less favored as higher temperatures are typically needed to dehydroxylate $Al_2O_3$. However, nondissociatie activation of $H_2O$ is the contributing one to form a formic acid intermediate.

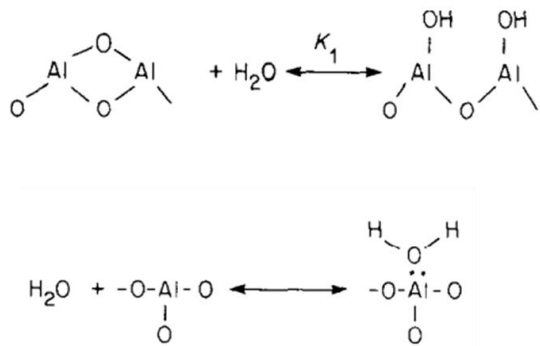

Figure 2. Two ways of $H_2O$ activation (Adapted from [12])

Formic acid-surface complex was proposed to be an intermediate from the reaction of metal activated CO and support activated water (Figure 3), and the exchange of formic acid from support sites onto metal sites was thought to be the rate limiting step. This step is crucial as on the Lewis acid site of support surface, absorbed formate would only decompose into CO and $H_2O$, while on the metal surface, it would decompose selectively into $CO_2$ and $H_2$ at a very high speed, as demonstrated by individual studies on formic acid decomposition.

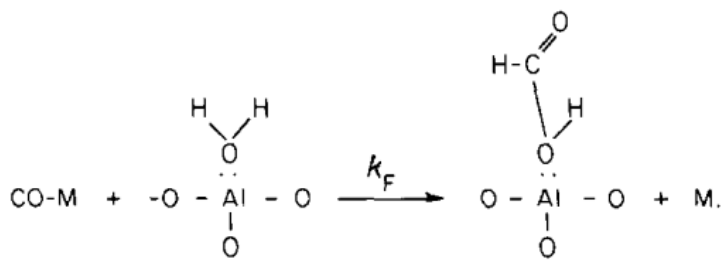

Figure 3. Formation of formic acid intermediate (Adapted from [12])

**Bifunctional reaction mechanism**

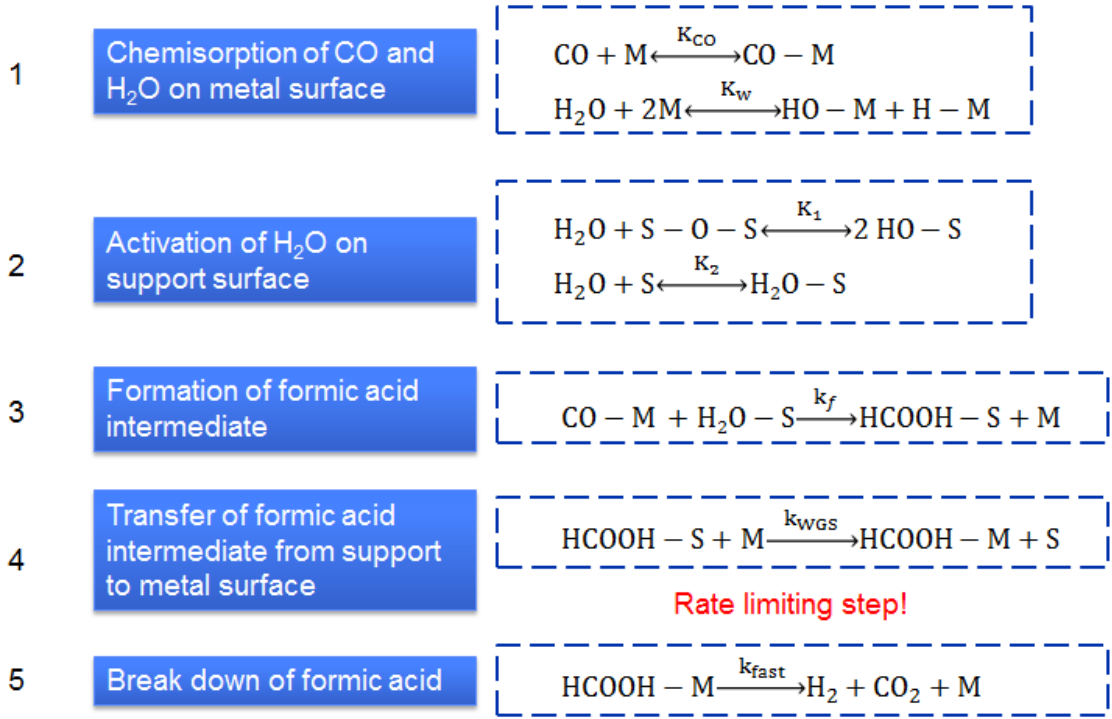

**Scheme 2. Summary of bifunctional reaction mechanism**

The bifunctional reaction mechanism in this study has been summarized in Scheme 2. The chemisorption of CO and $H_2O$ on metal surface, and $H_2O$ on support surface are the initiation steps happening simultaneously. And they are fast enough that all the species are in equilibrium. Thus, the Langmuir adsorption isotherm model can be readily used to solve the surface concentration of CO-M and $H_2O$-S.

The formation of formic acid intermediate and its further break down on metal surface are fast enough compared with the transfer of formic acid intermediate from support to metal, which was thought of as the rate limiting step.

By plugging in the Langmuir isotherm expressions for concentration of surface species, making binding assumptions of CO and $H_2O$ on metal, and further approximating using a mathematical relationship, the final expression for the rate of WGS can be boiled down to:

$$r = kP_{CO}^{X} P_{W}^{(1-X)/2}$$

**Eqn 1. Final expression of WGS reaction rate (Adapted from [12])**

Eqn 1 shows a very interesting connection between CO and water in terms of reaction order. For most of the metals investigated, they have an approximate reaction order of 0 for CO and of 0.5 for $H_2O$. This result is in great accordance with Eqn 1, suggesting the correctness of the proposed reaction mechanism.

**Validation of the proposed mechanism**

To further validate the proposed bifunctional mechanism, this paper plotted the calculated reaction order of $H_2O$ from the expression (1-X)/2, where X is the experimentally determined CO order of reaction, against the experimental determined $H_2O$ reaction order. If Eqn 1 is correct, then most of the data points should be on or around the line Y=X in this plot. Figure 4 shows such a nice agreement between calculated and experimental water reaction order for all metals studied, strongly supporting the hypothesis of a formic acid intermediate and a bifunctional reaction mechanism.

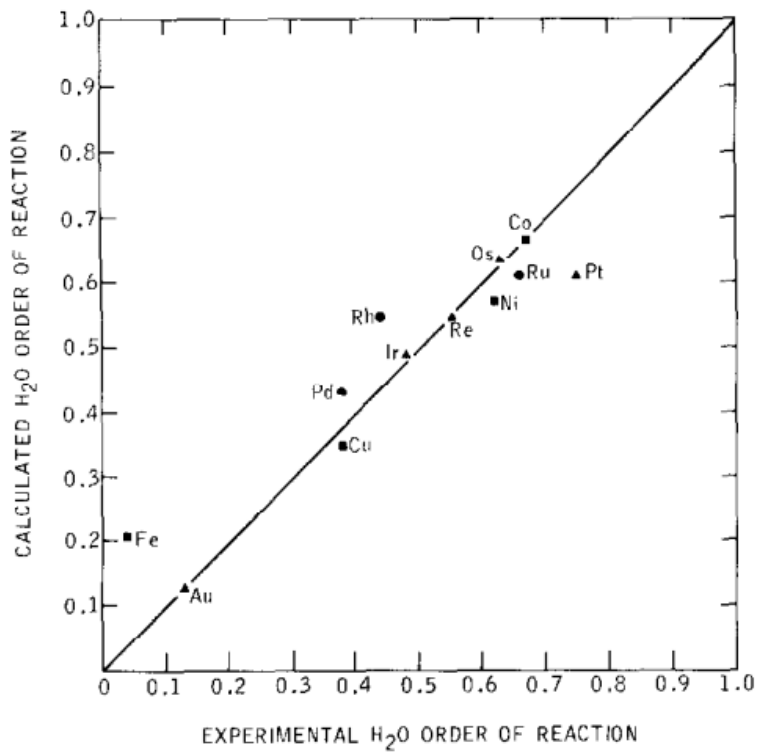

**Figure 4. Plot of calculated H₂O order of reaction vs. experimental order of reaction (Adapted from [12])**

**Conclusion**

The authors found out that the turn over numbers of various $Al_2O_3$ supported transition metals decreased in the trend of Cu, Re, Co, Ru, Ni, Pt, Os, Au, Fe, Pd, Rh, and Ir. The activity trend corresponds nicely to the observed volcano shaped correlation between catalytic activities and respective CO adsorption heat, indicative of CO activation on metal surface. The authors also observed that the turn over number of $Pt/Al_2O_3$ was one order of magnitude higher than that of $Pt/SiO_2$, indicating a strong support effect, which the authors ascribed to the activation of $H_2O$. A bifunctional reaction mechanism was proposed assuming formic acid is the intermediate and its transfer from support to metal surface is the rate limiting step. The proposed mechanism was validated by the experimental data, showing that the partial pressure dependencies for CO and H$_2$O are close to $0^{th}$ and $0.5^{th}$ order for most metals, respectively.